\documentclass[5p, times]{elsarticle}

\usepackage[normalem]{ulem}
\usepackage{lineno,hyperref}
\usepackage{graphicx}
\usepackage{kotex}
\usepackage{multirow}
\usepackage{framed}
\usepackage{subcaption}
\usepackage{lipsum}
\usepackage{comment}
\usepackage{tablefootnote}
\usepackage{mathtools}
\usepackage{amsmath,amssymb,amsfonts, amsthm}
\usepackage{multirow}
\usepackage{enumitem} 
\usepackage{amsmath}
\usepackage{threeparttable}
\usepackage{subcaption}
\usepackage{footnote}
\usepackage{caption}
\usepackage{tikz}
\usepackage{stfloats} 
\usepackage{thmtools}
\usepackage[linesnumbered,boxruled,ruled,noend]{algorithm2e}

\newtheorem{definition}{Definition}
\newtheorem{property}{Property}

\newtheorem{example}{Example}

\newcommand{\spara}[1]{\smallskip\noindent{\bf #1}}

\begin{document}

\makeatletter
\def\ps@pprintTitle{%
  \let\@oddhead\@empty
  \let\@evenhead\@empty
  \let\@oddfoot\@empty
  \let\@evenfoot\@oddfoot
}
\makeatother

\begin{frontmatter}

\title{Exploring Cohesive Subgraphs in Hypergraphs: The $(k,g)$-core Approach}


\author[unist]{Dahee Kim}
\ead{dahee@unist.ac.kr}

\author[unist]{Junghoon Kim\corref{mycorrespondingauthor}}
\cortext[mycorrespondingauthor]{Corresponding authors}
\ead{junghoon.kim@unist.ac.kr}

\author[cnu]{Sungsu Lim}
\ead{sungsu@cnu.ac.kr}

\author[knu]{Hyun Ji Jeong}
\ead{hjjeong@kongju.ac.kr}

\address[unist]{Ulsan National Institute of Science \& Technology, 44191, South Korea}
\address[cnu]{Chungnam National University, 34134, South Korea}
\address[knu]{Kongju National University, 31080, South Korea}

\begin{abstract}
Identifying cohesive subgraphs in hypergraphs is a fundamental problem that has received recent attention in data mining and engineering fields. Existing approaches mainly focus on a strongly induced subhypergraph or edge cardinality, overlooking the importance of the frequency of co-occurrence. In this paper, we propose a new cohesive subgraph named $(k,g)$-core, which considers both neighbour and co-occurrence simultaneously. The $(k,g)$-core has various applications including recommendation system, network analysis, and fraud detection. To the best of our knowledge, this is the first work to combine these factors. We extend an existing efficient algorithm to find solutions for $(k,g)$-core. Finally, we conduct extensive experimental studies that demonstrate the efficiency and effectiveness of our proposed algorithm.

\end{abstract}


\end{frontmatter}


\section{Introduction}
Complex systems such as social networks, biological networks, and market transaction systems are becoming increasingly intricate. The ability to accurately model and analyse these systems is important for understanding their behaviour and predicting their evolution. Traditional graph theory~\cite{west2001introduction}, which has been widely used to represent relationships between entities in a system, often falls short in capturing the high-order relationships in complex systems~\cite{berge1984hypergraphs}. 
Hypergraphs~\cite{bretto2013hypergraph}, which generalise traditional graph theory, offer a more flexible framework by allowing edges to connect any number of nodes, unlike traditional graphs where edges connect only two nodes. This characteristic makes hypergraphs particularly suitable for capturing multifaceted relationships and interactions in real-world networks, ranging from social networks to biological systems and beyond.

Real-world networks exhibit complex interactions 
that go beyond pairwise connections. For example, co-authorship networks~\cite{lung2018hypergraph}, co-purchase networks~\cite{xia2021self}, and location-tagged social networks~\cite{kumar2020hpra} have complex relationships that cannot be adequately represented by traditional graphs. Hypergraphs provide more expressive representations for such networks, enabling a deeper understanding of underlying structures and interactions~\cite{berge1984hypergraphs}.

Despite the increasing popularity and wide-ranging applications of mining hypergraphs~\cite{lee2022mining}, understanding their structural properties remains a challenging task. One of the key aspects of this challenge is to find cohesive subgraphs within a hypergraph, which are more densely connected internally than with the rest of the graph. 
In general, identifying the cohesive structure is important for understanding the overall structure and function of the graph, as they often represent key communities or roles~\cite{malliaros2020core}. In addition, it provides insights into the robustness of networks~\cite{giatsidis2011evaluating}, aiding in network analysis~\cite{adiga2013robust} and identifying influential nodes~\cite{kitsak2010identification}, and revealing query-centric communities~\cite{fang2020survey}.

\begin{table*}[t]
\centering
\footnotesize
\caption{Model comparison}
\begin{tabular}{c|cccccc}
\hline
          & \multicolumn{1}{c|}{\textbf{nbr-$k$-core}}                           & \multicolumn{1}{c|}{\textbf{$(k,d)$-core}}                               & \multicolumn{1}{c|}{\textbf{$(k,q)$-core}}            & \multicolumn{1}{c|}{\textbf{Clique-core}} & \multicolumn{1}{c|}{\colorbox{white}{\textbf{$(k,g)$-core}} }              & \textbf{$(\alpha, \beta)$-core}     \\ \hline \hline
objective & \multicolumn{6}{c}{maximize subgraph size}                                                                                                                                                                      \\ \hline
1st param & \multicolumn{2}{c|}{strongly induced. subgraph}               & \multicolumn{1}{c|}{deg. const.}        & \multicolumn{2}{c|}{neighbour size const.}                                        & deg. const.        \\ \hline
2nd param & \multicolumn{1}{c|}{-}     & \multicolumn{1}{c|}{deg. const.} & \multicolumn{1}{c|}{cardinality const.} & \multicolumn{1}{c|}{-}                 & \multicolumn{1}{c|}{co-occur. const.} & cardinality const. \\ \hline \hline
\end{tabular}
\end{table*}

Recently, several cohesive models have been proposed for hypergraphs. The first approach involves converting the hypergraph into a clique-like graph, then applying $k$-core algorithm~\cite{nbrk, seidman1983network}. $k$-core is a maximal set of nodes of which every node has at least $k$ neighbour nodes in the $k$-core. We refer to this approach as the clique-core.
Lee et al.~\cite{kqcore} introduced the $(k,q)$-core, a maximal subgraph in which each node has at least $k$ degree, and each hyperedge contains at least $q$ nodes. With the similar period, the nbr-$k$-core was proposed by Arafat et al.~\cite{nbrk}. The nbr-$k$-core is the maximal strongly induced connected subhypergraph~\cite{bahmanian2015connection, graham1996handbook} $H$ such that every node has at least $k$ neighbours in $H$. Note that the strongly induced subhypergraph indicates that every node in the hyperedge belongs to the subhypergraph if and only if all nodes of the hyperedge exist.  Arafat et al.~\cite{nbrk} also formulated the $(k,d)$-core, a maximal strongly induced connected subhypergraph in which every node has at least $k$ neighbours and degree (number of hyperedges incident on a node) is greater than or equal to $d$ in strongly induced subhypergraph.

While these proposed models have proven useful, they may not capture all aspects of cohesion in hypergraphs. Especially, hyperedges in hypergraphs have some distinct characteristics: 
(1) \textit{hyperedge nesting}: Hyperedges often nest, with one's vertex set being a subset of another's, representing hierarchical or containment relationships. For instance, in biological networks, smaller modules may nest within larger ones, and in transaction networks, certain products may repeatedly co-occur in transactions; 
(2) \textit{nodes belong to multiple hyperedges}: Nodes can engage in multiple relationships, each depicted by a hyperedge. In social networks, an individual may be associated with different communities or groups through distinct hyperedges.

However, existing models do not mainly consider these characteristics when designing cohesive subgraph models. To effectively address the above issues, we incorporate the frequency of co-occurrence of nodes within hyperedges, which can be a crucial aspect of cohesion in many systems.
Taking into account the frequency with which entities co-occur has proven to be effective in domains~\cite{kim2022ocsm} such as recommendation systems. This is particularly effective in the context of user-based collaborative filtering~\cite{breese2013empirical, Mohan}. In user-based collaborative filtering, it is important to find users who have similar preferences or tastes to a given user. This is because fundamental principle of user-based collaborative filtering is to recommend items based on the choices of similar users. Cohesive subgraphs of hypergraphs based on the frequency of co-occurrence among users help to define target similar users for capturing collaborative signals. 

Therefore, we propose a new concept $(k,g)$-core by extending traditional $k$-core~\cite{seidman1983network} by incorporating the occurrence of the cohesive subgraphs. The $(k,g)$-core is a maximal subgraph in which each node has at least $k$ neighbours which appear in at least $g$ hyperedges together in the $(k,g)$-core. 
Note that $(k,g)$-core can be solved by converting a clique-based graph. However, it is not desired since this step inflates the size of the problem~\cite{huang2015scalable}. 
In the following, we point out the motivating applications of the $(k,g)$-core.

\spara{Applications.} 
(1) Fraudster Detection: Within a social commerce service, user purchasing data can be organised into a hypergraph, where edges represent groups of items bought by users. Fraudsters often exhibit similar behaviours and have connections through shared product purchases~\cite{kim2022abc}. For example, a seller may recruit individuals to inflate review ratings for their mobile devices, resulting in fraudulent activity. By identifying shared characteristics among users, it becomes possible to cluster these fraudsters within the hypergraph network; (2) Biomedical Systems: The $(k,g)$-core model finds extensive application in modeling biological systems, such as protein-protein interaction networks and gene regulatory networks~\cite{feng2021hypergraph}. Leveraging this model helps to identify highly interconnected substructures within these networks, revealing crucial protein complexes or functional modules that play key roles in biological processes; and 
(3) Recommendation Systems:  The $(k,g)$-core model offers valuable insights in recommendation systems by finding user clusters with similar preferences. By detecting $(k,g)$-core subgraphs in the hypergraph representing user-item interactions, personalised recommendations can be generated, capitalising on the co-occurrence patterns of items within these subgraphs.


\section{RELATED WORK}
The study of hypergraphs has seen a surge of interest in recent years, with numerous methods proposed based on different criteria~\cite{luo2021hypercore, wang2022efficient}. In this section, we review the most relevant previous study. 

\spara{$(k,q)$-core~\cite{kqcore}}. One of the models for finding cohesive subgraphs by considering hyperedges and nodes together in hypergraphs is the $(k,q)$-core, proposed by Lee et al. This model defines a $(k,q)$-core as the largest subgraph in which each node has at least $k$ degree, and each hyperedge contains at least $q$ nodes. Even if this model has various applications, providing valuable insights into the structure of hypergraphs, it does not consider the frequency of co-occurrence of nodes within hyperedges, which can be a crucial aspect of cohesion in many systems.

\spara{nbr-$k$-core~\cite{nbrk}}. 
nbr-$k$-core is proposed by Arafat et al. It is the maximal strongly induced subhypergraph~\cite{bahmanian2015connection, graham1996handbook} such that every node has at least $k$ neighbours. The concept strongly induced subhypergraph implies that for each node within a hyperedge, it must be included in the subhypergraph if and only if all other nodes within the same hyperedge are also present. Note that utilising a strongly induced subhypergraph may have some issues. A hyperedge can be present in its entirety or not at all in a subhypergraph due to its definition. When a node is included in a large hyperedge, a size of the strongly induced subhypergraph can be very large, leading to difficulties in analysing an extracting meaningful information.

\spara{$(k,d)$-core~\cite{nbrk}}. 
Arafat et al.~\cite{nbrk} observed the problem that the nbr-$k$-core usually return a very large cohesive subgraph as a result if it is included in a large-sized hyperedge. To address this issue, they proposed a more comprehensive cohesive subgraph model, $(k,d)$-core, by considering neighbourhood and degree constraints simultaneously. The $(k,d)$-core is defined the maximal subhypergraph in which every node has at least $k$ neighbours and degree of every node is at least $d$ in strongly induced subhypergraph. However, this approach could not capture the strength of the neighbours due to the lack of consideration for co-occurrence.  

\spara{Clique-core~\cite{batagelj2011fast}}. The Clique-core is the same with traditional $k$-core by converting a hypergraph as clique-structured graph. After preprocessing, it aims to find a maximal set of nodes of which every node has at least $k$ neighbour nodes in the clique graph. This clique-based approach is known as inflating the problem-size~\cite{huang2015scalable}.

\spara{$(\alpha, \beta)$-core~\cite{alphabeta}}. The $(\alpha, \beta)$-core is another model widely used in the study of bipartite networks. It aims to find a set of nodes in the bipartite graph where each side has at least $\alpha$ and $\beta$ bipartite edges, respectively. A hypergraph can be converted to a bipartite graph that consists of nodes as the first set of nodes, and the hyperedge as other node set and creates an edge between two node sets if the hyperedge in the second set involves a node in the first set. 
While this model has proven useful in bipartite networks, it cannot be directly utilised to find cohesive subgraphs in hypergraphs due to the lack of neighbour structure information. In addition, constructing a bipartite graph may suffer from size inflating problem~\cite{huang2015scalable}.

\section{PROBLEM STATEMENT}
A hypergraph network can be modelled as a graph $G=(V,E)$ with nodes $V$ and edges $E$.  Following previous studies~\cite{nbrk, kqcore}, we consider that $G$ is undirected and unweighted. 
In hypergraph notation, we use the term \textit{degree} to represent the count of hyperedges incident to a particular node. Moreover, we employ the term \textit{neighbour} to refer to any two nodes that appear together in a hyperedge.

\begin{definition}
    (\underline{$(k,g)$-core}). Given a hypergraph $G$, $k\geq 1$ and $g\geq 1$, $(k,g)$-core is the maximal set of nodes in which each node has at least $k$ neighbours which appear in at least $g$ hyperedges together in an induced subhypergraph by the $(k,g)$-core.
\end{definition}

We next discuss two essential properties of $(k,g)$-core: \textit{uniqueness} and \textit{containment}. These properties are fundamental for understanding the behaviour and structure of the $(k,g)$-core.

\begin{figure}[t]
\centering
\includegraphics[width=0.99\linewidth]{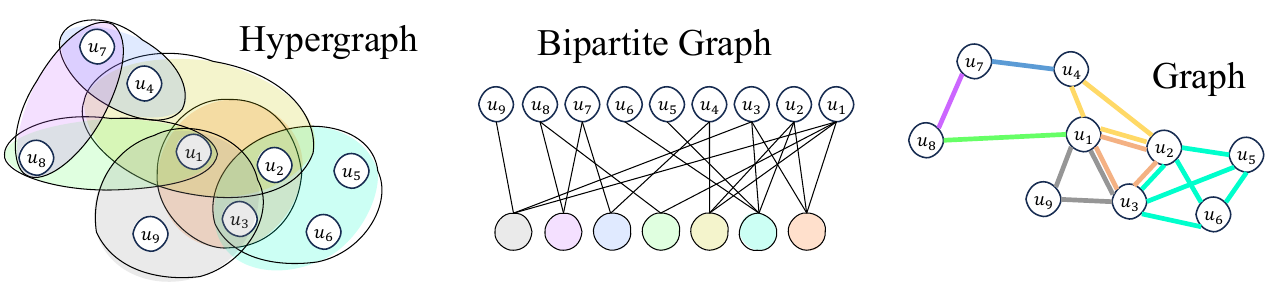}
\caption{Motivating example} 
\vspace*{-0.3cm} 
\label{fig:intro}
\end{figure}

\begin{property}
\underline{Uniqueness}: $(k,g)$-core is unique. 
\end{property}

\begin{proof}
$(k,g)$-core is unique due to the maximality constraint, as it represents the maximal set of nodes in which all nodes have at least $k$ neighbours which appear in at least $g$ hyperedges, achieved by iteratively removing nodes until satisfying the constraints. 
\end{proof}

\begin{property}
\underline{Containment}: $(k,g)$-core has hierarchical structure, i.e., $(k+1,g)$-core $\subseteq$ $(k, g)$-core and $(k,g+1)$-core $\subseteq$ $(k,g)$-core. 
\end{property}

\begin{proof}
The hierarchy structure of the $(k,g)$-core is a result of a gradual node removal process, ensuring that if a node belongs to the $(k,g)$-core, it also belongs to all higher-order cores due to the cumulative neighbour and co-occurrence constraints. satisfied by the remaining nodes.
\end{proof}

\begin{example}
Let us consider an example using a simple hypergraph with $9$ nodes and $7$ hyperedges, as illustrated in Figure~\ref{fig:intro}. This example demonstrates the distinct results and characteristics of different cohesive subgraph models.
\begin{itemize}[leftmargin=*]
    \item The $(k,q)$-core yields $\{u_1, u_2, u_3, u_4, u_7, u_8\}$ with $k=2$ and $q=2$. This is because the nodes $u_5$, $u_6$, and  $u_9$ are iteratively removed due to the degree constraint. The remaining graph then satisfies the $(k,q)$-core constraint. 
    \item The nbr-$k$-core returns the entire graph when $k$=2, and $\{u_2, u_3,$ $u_5, u_6\}$ when $k$=3 since there are hyperedges involving only $u_2$,$u_3$,$u_5$ and $u_6$, every node has 3 neighbours.
    \item The $(k,d)$-core returns the entire graph for $k=2$ and $d=1$, $\{u_2, u_3, u_5,u_6\}$ for $k=3$ and $d=1$. 
    When $k=2$ and $d=2$, it returns an empty set. We can observe that $u_1, u_2, u_3, u_4, u_7,$ and $u_8$ have two or more neighbours and belong to at least two hyperedges. Note that the $(k,d)$-core is a strongly induced subhypergraph. Thus, hyperedges such as $\{u_1, u_3, u_9\}$ and $\{u_2, u_3, u_5, u_6\}$ are not considered. 
    Therefore, the node $u_3$ must be removed. This removal process is repeated iteratively until no nodes remain in the resulting graph.
    \item Clique-core returns the entire graph for $k=2$, and returns $\{u_2, u_3, u_5$ $, u_6\}$ for $k=3$ due to the neighbour constraint. 
    \item The $(\alpha, \beta)$-core returns $\{u_1, u_2, u_3, u_4, u_7, u_8\}$ for $\alpha=2$ and $\beta=2$. Since every node is included in at least $2$ hyperedge, and a hyperedge has at least two nodes, $u_5$, $u_6$ and $u_9$ cannot be included because they belong to only a single hyperedge.
    \item Lastly, the $(k,g)$-core gives $\{u_1, u_2, u_3\}$ for $k=2$ and $g=2$. In this case, each node has at least $k$ neighbours and each node pair appears in at least $g$ edges together, signifying a cohesive structure within the hypergraph. 
\end{itemize}
\end{example}
\begin{algorithm}[t]
\footnotesize
\SetAlgoLined
\SetKwData{break}{break}
\SetKwData{return}{return}
\SetKwData{true}{true}
\SetKwData{false}{false}
\SetKwFunction{degree}{degree}
\KwIn{Hypergraph $G=(V,E)$, parameters $k$ and $g$}
\KwOut{The $(k,g)$-core of $G$}
$H \leftarrow V$ \tcp*{Initialise $H$ as the set of all nodes}
Initialise neighbour occur. map $NOM$\;
\ForEach{$v \in H$}{
    \ForEach{hyperedge $e$ that contains $v$}{
        \ForEach{$u \in e$}{
            \If{$u \neq v$}{
                Increment the occur. count of $u$ in $NOM[v]$;
            }
        }
    }
}
Keep neighbours in $NOM[v]$ for each $v\in V$ where the occur. count is $\geq g$\;
$changed \leftarrow \text{True}$\;
\While{$changed$}{
    $changed \leftarrow \text{False}$\;
    \For{$w \in H$}{
        \If{$|NOM[w]| < k$}{
            $H \leftarrow H \setminus \{w\}$\;
            Update the occur. map\;
            $changed \leftarrow \text{True}$\;
        }
    }
}
\Return{$H$}
\caption{Peeling Algorithm for $(k,g)$-core}
\label{alg:algorithm}
\end{algorithm}
\setlength{\textfloatsep}{5pt}
\section{PEELING ALGORITHM}\label{sec:algorithm}
In this section, we present the peeling algorithm to find $(k,g)$-core. We extend the existing peeling algorithm for the $k$-core computation~\cite{batagelj2003m}. Basically, it iteratively removes a set of nodes until satisfying the cohesiveness constraint. For each iteration, it finds nodes which do not satisfy the $(k,g)$-core criteria, and removes them from the hypergraph. The algorithm continues until no more nodes can be removed. Finally, the remaining subgraph can be the $(k,g)$-core. 

The pseudo description of the algorithm can be checked in Algorithm~\ref{alg:algorithm}. It starts by initialising $H$ as the set of all nodes in the hypergraph. It then initialises the neighbour occurrence map, $NOM$, for each node in $V$ with an empty dictionary data structure. The algorithm iterates over each node in $H$ and counts the occurrences of its neighbours on the hyperedges. After that, it keeps only the neighbours in $NOM[v]$ for each node $v$ where the occurrence count is $\geq g$.
The algorithm proceeds in a loop that continues until no more changes are made. Within the loop, it iterates over each node $w$ in $H$. If $NOM[w]$ does not contain at least $k$ nodes, indicating that node $w$ does not satisfy the $(k,g)$-core criteria, it is removed from $H$, and the occurrence map is updated accordingly. This process continues until no more nodes can be removed. Finally, it returns the nodes $H$ as a result. 

This peeling method is efficient to find the $(k,g)$-core in a hypergraph. The time complexity of the peeling algorithm is $O(|V|^2 |E|)$. In the worst-case, the algorithm iterates over each node and its corresponding hyperedges to construct the occurrence map. As each hyperedge can contain up to $|V|$ nodes, constructing the occurrence map for each node takes $O(|V||E|)$ time, resulting in a time complexity of $O(|V|^2 |E|)$ for this step. 
During the iterative removal procedure, the algorithm performs a maximum of $|V|$ iterations, and each removal operation has a time complexity of $O(|V|)$. As a result, the overall time complexity is $O(|V|^2|E|)$.
However, it's important to note that the algorithm may terminate earlier as nodes can be removed together. 
\section{EXPERIMENTS}

To validate effectiveness of the proposed $(k,g)$-core and efficiency of the peeling algorithm, we conducted extensive experiments on real-world hypergraphs. 

\spara{Experimental setup.} We implemented the $(k,g)$-core model in Python using the NetworkX library~\cite{hagberg2008exploring}. The experiments were run on a Linux machine with Intel Xeon 6248R and 256GB of RAM. 
%
\begin{table}[t]
\footnotesize 
\centering
\caption{Real-world dataset}
\label{tab:data}
\begin{tabular}{c||c|c|c|c}
\hline
       Dataset  & \textbf{$|V|$}         & \textbf{$|E|$}         &\textbf{Avg. nbr size} & \textbf{Avg. edge card.} \\ \hline \hline

\textbf{Contact}  & 242       & 12,704    & 68.74            & 2.42                   \\ \hline
\textbf{Congress} & 1,718     & 83,105    & 494.68           & 8.81                   \\ \hline
\textbf{Enron}    & 4,423     & 5,734     & 25.35            & 5.25                   \\ \hline
\textbf{Meetup}   & 24,115    & 11,027    & 65.27            & 10.3                   \\ \hline
\textbf{DBLP}     & 1,836,596 & 2,170,260 & 9.05             & 3.43                   \\ \hline
\textbf{Aminer} & 27,850,748     & 17,120,546    & 8.39           & 3.77                   \\ \hline \hline
\end{tabular}
\end{table}
%

\spara{Dataset.} Table~\ref{tab:data} provides the essential statistics of six real-world datasets. These datasets are publicly available and can be accessed from the sources mentioned in the references~\cite{nbrk, benson2018simplicial}. In the table, the term `nbr' denotes a neighbour, and `card' indicates cardinality.

\spara{Evaluation measure.} We evaluated the performance of the $(k,g)$-core model by checking the number of nodes and the running time.

\begin{figure}[ht]
    \centering
    \begin{subfigure}[b]{0.99\linewidth}
        \includegraphics[width=\linewidth]{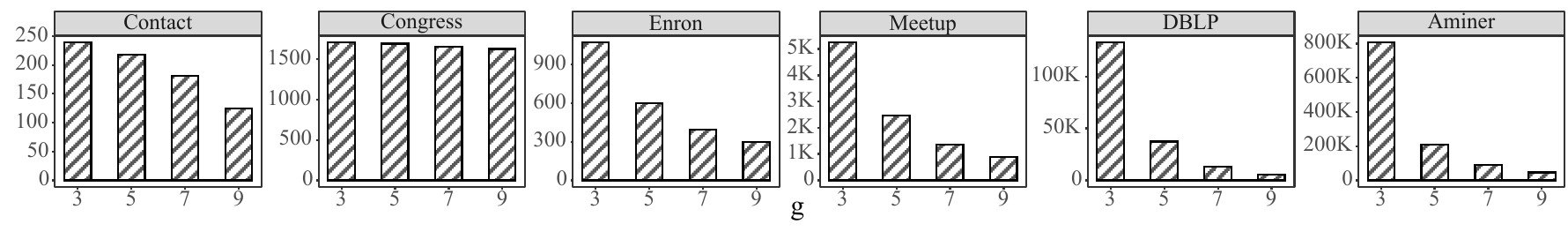}
        \caption{Fixing $k=3$}
        \label{fig:var_g_nodes}
    \end{subfigure}
    \begin{subfigure}[b]{0.99\linewidth}
        \includegraphics[width=\linewidth]{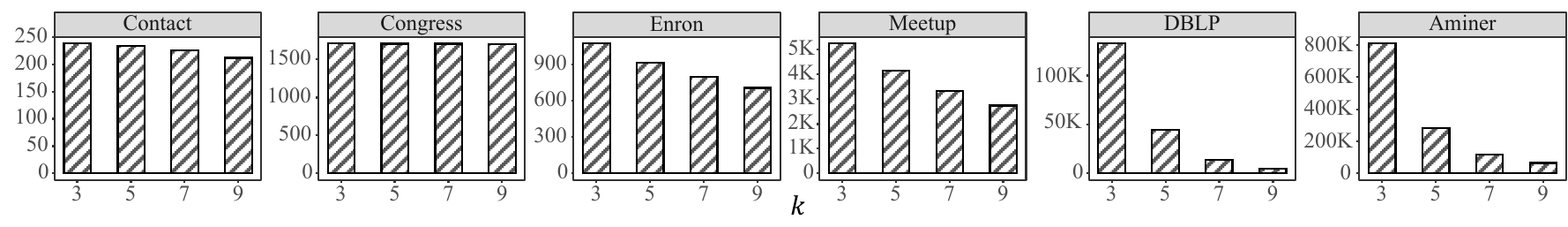}
        \caption{Fixing $g=3$}
        \label{fig:var_k_nodes}
    \end{subfigure}
    \caption{Comparison of number of nodes}
    \label{fig:nodes_comparison}
\end{figure}

\spara{Performance of $(k,g)$-core.} We varied the user parameters to analyse the behaviour of the $(k,g)$-core.
In Figure~\ref{fig:var_g_nodes}, we fix the value $k$ as $3$ and vary $g$ as $3$, $5$, $7$, and $9$. The experimental results report the number of nodes in the $(k,g)$-core for each $g$ value.
The results show that an increase in the value of $g$ also leads to a decrease in the number of nodes within the $(k,g)$-core. 
Similarly, in Figure~\ref{fig:var_k_nodes}, the value of $g$ is fixed as $3$, while $k$ is varied as $3$, $5$, $7$, and $9$. 
It is observed that when $k$ increases, the number of nodes in the $(k,g)$-core decreases. Both results indicate that a higher value of $k$ and $g$ results in more densely connected.

\begin{figure}[h]
    \centering
    \begin{subfigure}[b]{0.99\linewidth}
        \includegraphics[width=\linewidth]{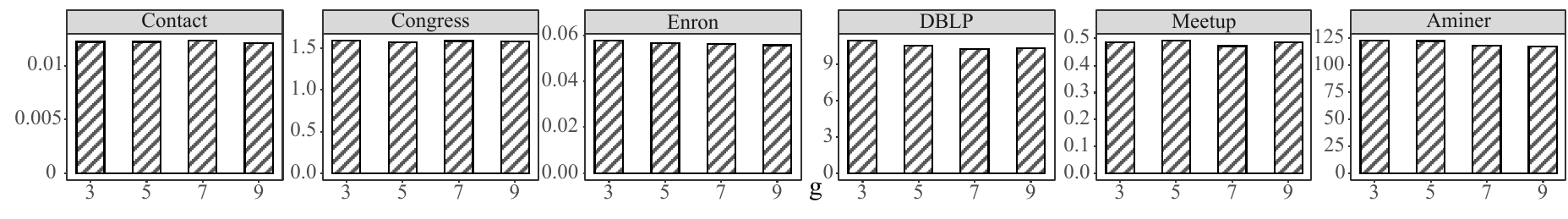}
        \caption{Fixing $k=3$}
        \label{fig:var_g_times}
    \end{subfigure}
    \begin{subfigure}[b]{0.99\linewidth}
        \includegraphics[width=\linewidth]{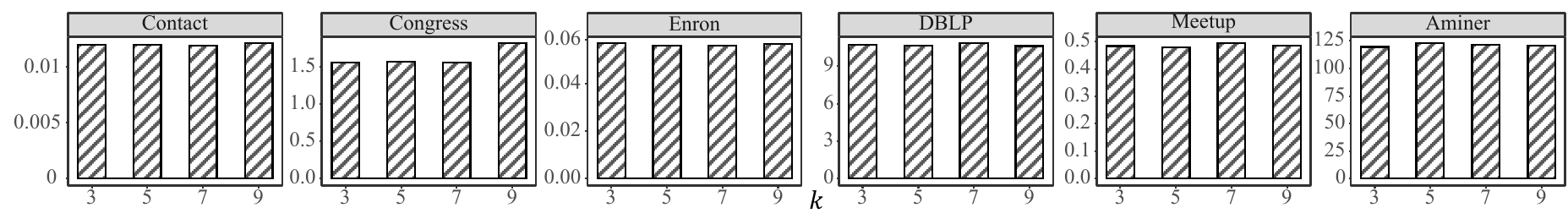}
        \caption{Fixing $g=3$}
        \label{fig:var_k_times}
    \end{subfigure}
    \caption{Comparison of running time}        
   \label{fig:times_comparison}
\end{figure}

Next, we focus on analysing the running time of the algorithm while keeping the parameters fixed. 
Figure~\ref{fig:times_comparison} presents the results obtained by varying the values of $k$ and $g$. The experimental results indicate that increasing $k$ and $g$ does not have a significant impact on the running time. This implies that the running time of the $(k,g)$-core is not primarily determined by the user-defined parameters $k$ and $g$. Instead, the majority of the computational complexity arises from calculating the occurrence of the neighbours as we have discussed in Section~\ref{sec:algorithm}.

\begin{figure}[h]
\centering
\includegraphics[width=0.6\linewidth]{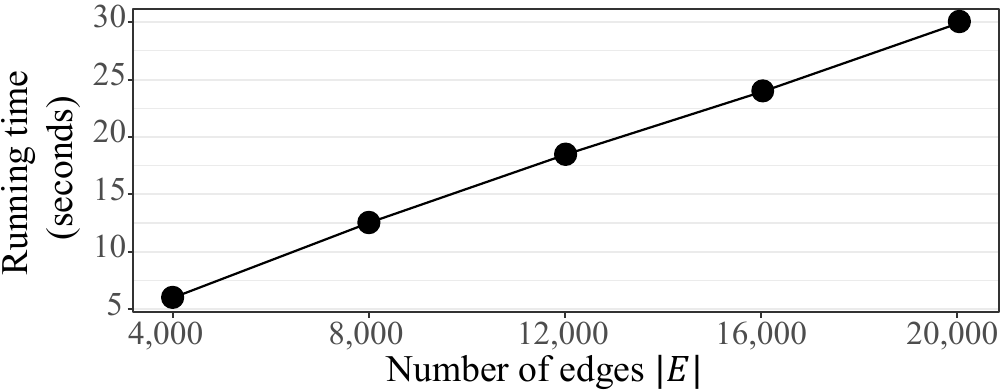}
\caption{Scalability test} 
\label{fig:scalability}
\end{figure}

\spara{Scalability test.} 
To evaluate scalability of the algorithm, we conducted a scalability test using $k$-uniform hypergraph generation models~\cite{feng2019hypergraph}. With a fixed value of $k$ ($100$) and a set number of nodes ($10,000$), we varied the number of hyperedges from 4,000 to $20,000$ to observe the impact on running time. By converting the hypergraph into a bipartite network, we obtained a bipartite graph with $30,000$ nodes and  $2$ million bipartite edges when the number of hyperedges was $20,000$. Figure~\ref{fig:scalability} shows the experimental result. It reveals a linear increase in running time when the number of hyperedges increases, indicating that the algorithm scales efficiently with larger hypergraphs. This scalability test demonstrates the algorithm's capability to handle large-sized hypergraphs.

\section{CONCLUSION}

In this paper, we introduce the $(k,g)$-core model for cohesive subgraph discovery in hypergraphs, an extension of the existing $k$-core model that takes into account node co-occurrence within hyperedges. We also propose a peeling algorithm that effectively identifies the $(k,g)$-core by iteratively removing nodes that do not satisfy the specified criteria. Experimental evaluations on six real-world networks demonstrate the characteristics of proposed $(k,g)$-core. In future work, we plan to recommend appropriate values for $k$ and $g$ to identify meaningful cohesive subgraphs, thereby eliminating the need for user-specified parameters.

\bibliography{000_bib.bib}

\end{document}